\documentclass[sigconf]{acmart}
\usepackage{amsmath,bm}

\usepackage{enumitem}

\usepackage{soul}
\setul{0.5ex}{0.1ex}
\setulcolor{gray}

\usepackage{subcaption}
\usepackage{dblfloatfix}  

\usepackage{xspace}

\newcommand{\toolbox}{\textsc{Tiger}\xspace{}}

\DeclareMathAlphabet{\altmathcal}{OMS}{cmsy}{m}{n}

\usepackage{listings}
\usepackage{xcolor}

\definecolor{codegreen}{rgb}{0,0.6,0}
\definecolor{codegray}{rgb}{0.5,0.5,0.5}
\definecolor{codepurple}{rgb}{0.58,0,0.82}
\definecolor{backcolour}{rgb}{0.95,0.95,0.92}

\lstdefinestyle{mystyle}{
    commentstyle=\color{codegreen},
    keywordstyle=\color{magenta},
    numberstyle=\tiny\color{codegray},
    stringstyle=\color{codepurple},
    basicstyle=\ttfamily\footnotesize,
    breakatwhitespace=false,         
    breaklines=true,                 
    captionpos=b,                    
    keepspaces=true,                 
    numbers=left,                    
    numbersep=5pt,                  
    showspaces=false,                
    showstringspaces=false,
    showtabs=false,                  
    tabsize=2,
    xleftmargin=0.25cm,
    frame=tb,
    aboveskip=0.5cm
}

\usepackage{xcolor}
\usepackage{multirow}

\usepackage{rotating}
\usepackage{colortbl}
\usepackage{tcolorbox}
\usepackage{tikz}
\usepackage{arydshln}
\usepackage{makecell}

\usepackage{pifont}

\usepackage{caption}
\usepackage{array}
\newcolumntype{L}{>{\centering\arraybackslash}m{6cm}}
\newcolumntype{L}[1]{>{\raggedright\let\newline\\\arraybackslash\hspace{0pt}}m{#1}}
\newcolumntype{C}[1]{>{\centering\let\newline\\\arraybackslash\hspace{0pt}}m{#1}}
\newcolumntype{R}[1]{>{\raggedleft\let\newline\\\arraybackslash\hspace{0pt}}m{#1}}
\newcolumntype{P}[1]{>{\raggedright}p{#1}}

\usepackage{booktabs}

\definecolor{why}{HTML}{EE220C}
\definecolor{who}{HTML}{00A2FF}
\definecolor{what}{HTML}{1DB100}
\definecolor{how}{HTML}{F8BA00}
\definecolor{when}{HTML}{CB297B}
\definecolor{where}{HTML}{00A89D}

\definecolor{white}{HTML}{FFFFFF}
\definecolor{lightgray}{HTML}{F3F3F3}
\definecolor{black}{HTML}{000000}
\definecolor{tabletagcolor}{HTML}{BDBDBD}

\definecolor{cell}{HTML}{B0BEC5}
\definecolor{rowbackground}{HTML}{F0F2F4}

\setcopyright{acmcopyright}
\copyrightyear{2021}
\acmYear{2021}
\acmDOI{10.1145/1122445.1122456}

\acmPrice{15.00}
\acmISBN{978-1-4503-XXXX-X/18/06}

\begin{document}


\title{Evaluating Graph Vulnerability and Robustness using TIGER}



\author{Scott Freitas}
\email{safreita@gatech.edu}
\affiliation{%
  \institution{Georgia Institute of Technology}
  \city{Atlanta}
  \state{Georgia}
  \country{USA} 
  \postcode{30308}
}

\author{Diyi Yang}
\email{dyang888@gatech.edu}
\affiliation{%
  \institution{Georgia Institute of Technology}
  \city{Atlanta}
  \state{Georgia}
  \country{USA}
  \postcode{30308}
}

\author{Srijan Kumar}
\email{srijan@gatech.edu}
\affiliation{%
  \institution{Georgia Institute of Technology}
  \city{Atlanta}
  \state{Georgia}
  \country{USA}
  \postcode{30308}
}

\author{Hanghang Tong}
\email{htong@illinois.edu}
\affiliation{%
  \institution{University of Illinois at Urbana-Champaign}
  \city{Urbana}
  \state{Illinois}
  \country{USA}
  \postcode{61801}
}

\author{Duen Horng Chau}
\email{polo@gatech.edu}
\affiliation{%
  \institution{Georgia Institute of Technology}
  \city{Atlanta}
  \state{Georgia}
  \country{USA}
  \postcode{30308}
}

\renewcommand{\shortauthors}{Freitas et al.}

\begin{abstract}
Network robustness plays a crucial role in our understanding of complex interconnected systems such as transportation, communication, and computer networks.
While significant research has been conducted in the area of network robustness, no comprehensive open-source toolbox currently exists to assist researchers and practitioners in this important topic.
This lack of available tools hinders reproducibility and examination of existing work, development of new research, and dissemination of new ideas.
We contribute \toolbox{}, an open-sourced Python toolbox to address these challenges.
\toolbox{} contains 22 graph robustness measures with both original and fast approximate versions;
17 failure and attack strategies; 15 heuristic and optimization-based defense techniques; and 4 simulation tools.
By democratizing the tools required to study network robustness, our goal is to assist researchers and practitioners in analyzing their own networks; and facilitate the development of new research in the field.
\toolbox{} has been integrated into 
the Nvidia Data Science Teaching Kit available to educators across the world; 
and Georgia Tech's Data and Visual Analytics class with over 1,000 students.
\toolbox{} is open sourced at: \url{https://github.com/safreita1/TIGER}. 
\end{abstract}



\maketitle

\section{Introduction}
\begin{figure*}[t]
    \centering
    \includegraphics[width=0.9\textwidth]{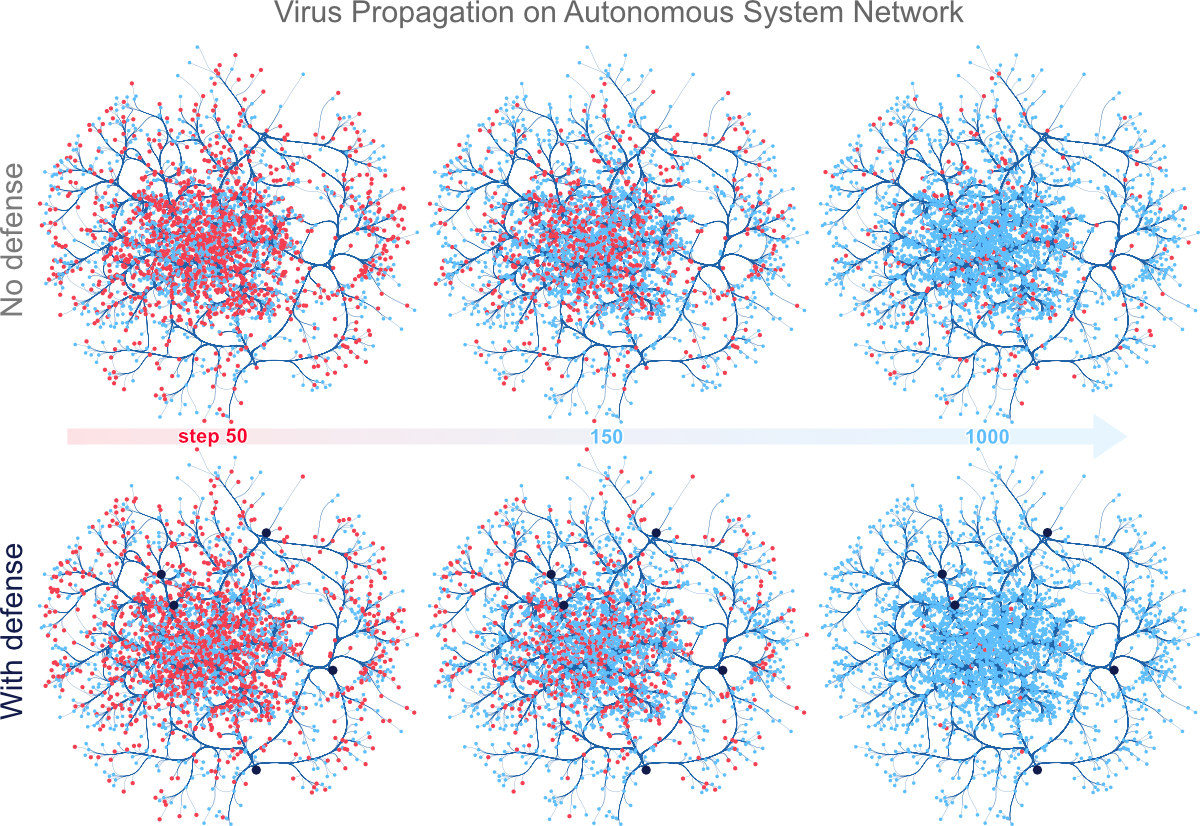}
    \caption{\toolbox{} provides a number of important tools for graph vulnerability and robustness research, including graph robustness measures, attack strategies, defense techniques and simulation models. 
    Here, a \toolbox{} user is visualizing a computer virus simulation that follows the SIS infection model (effective strength $s=3.21$) on the \textit{Oregon-1 Autonomous System} network~\cite{leskovec2005graphs}.
    Top: without any defense, the virus remains endemic.
    Bottom: defending only 5 nodes with  \textit{Netshield}~\cite{tong2010vulnerability}, the number of infected entities is reduced to nearly zero. 
}
    \label{fig:virus_simulation}
\end{figure*}

Through analyzing and understanding the robustness of networks we can: 
(1) quantify network vulnerability and robustness, 
(2) augment a network's structure to resist attacks and recover from failure, and
(3) control the dissemination of entities on the network (e.g., viruses, propaganda). 
Consider the impactful scenario where a virus penetrates one or more machines in an enterprise network.
Once infected, the virus laterally spreads to susceptible machines in the network, resulting in system-wide failures, data corruption and exfiltration of trade secrets and intellectual property.
This scenario is commonly modeled as a dissemination of entities problem using an epidemiological susceptible-infected-susceptible (SIS) model, where each machine is in either one of two states---infected or susceptible (see Figure~\ref{fig:virus_simulation}).
How quickly a virus spreads across a network is known as the network's \textbf{vulnerability}, and is defined as a \textit{measure of susceptibility to the dissemination of entities across the network}~\cite{tong2010vulnerability}.
A natural counterpart to network vulnerability is \textbf{robustness}, defined as a {\textit{measure of a network's ability to continue functioning when part of the network is naturally damaged or targeted for attack}}~\cite{ellens2013graph,chan2016optimizing,beygelzimer2005improving}

\vspace{0.1cm}
\noindent\textbf{Challenges for robustness and vulnerability research.} 
Network robustness has a rich and diverse background spanning numerous fields of engineering and science~\cite{klein1993resistance,beygelzimer2005improving,tong2010vulnerability,krishnamoorthy1987fault,freitas2020d2m}.
Unfortunately, this cross-disciplinary nature comes with significant challenges---resulting in slow dissemination of ideas, leading to missed innovation opportunities.
We believe a unified and easy-to-use software framework is key to standardizing the study of network robustness, helping accelerate reproducible research and dissemination of ideas.

\vspace{0.1cm}
\noindent\textbf{\toolbox{} design and implementation.} 
We present \toolbox{}, an open-sourced Python \textit{\textbf{T}oolbox for evaluat\textbf{I}ng \textbf{G}raph vuln\textbf{E}rability and \textbf{R}obustness}. 
Through \toolbox{}, our goal is to catalyze network robustness research, promote reproducibility and amplify the reach of novel ideas.
In designing \toolbox{}, we consider multiple complex implementation decisions, including: 
(1) the criterion for inclusion in the toolbox; 
(2) identifying and synthesizing a set of core robustivity features needed by the community; and 
(3) the design and implementation of the framework itself.
We address the \textit{inclusion criterion} by conducting a careful analysis of influential and representative papers (e.g., \cite{chan2016optimizing,tong2010vulnerability,malliaros2012fast,holme2002attack,beygelzimer2005improving}) across top journals and conferences from the relevant domains (e.g., ICDM, SDM, Physica A, DMKD, Physical Review E), many of which we will discuss in detail in this paper.
We also include papers posted on arXiv, as many cutting-edge papers are first released there.

Based on our analysis, we identify and include papers 
that tackle one or more of the following fundamental tasks~\cite{ellens2013graph,holme2002attack,beygelzimer2005improving}:
(1) measuring network robustness and vulnerability;
(2) understanding network failure and attack mechanisms; 
(3) developing defensive techniques;
and (4) creating simulation tools to model processes.
From these papers, we select and implement a total of 44 attacks, defenses and robustness measures, along with 4 simulation tools in which they can be used.
Due to a vibrant and growing community of users, we develop \toolbox{} in Python 3, leveraging key libraries, such as NetworkX, SciPy, Numpy and Matplotlib.
While excellent alternative network analysis tools exist~\cite{hagberg2008exploring,azizi2020epidemics,rossetti2018ndlib,klise2018overview,kang2018x,korkali2017reducing,freitas2017rapid,bastian2009gephi}, many of them are domain specific (e.g., EoN~\cite{azizi2020epidemics}, WNTR~\cite{klise2018overview}) or do not provide direct support for network robustness analysis (e.g., NetworkX~\cite{hagberg2008exploring}, Gephi~\cite{bastian2009gephi}).
In contrast, \toolbox{} complements existing tools while providing key missing network robustness components.

\subsection{Contributions}

\noindent\textbf{1. TIGER.}
We present \toolbox{}, the first open-sourced Python toolbox for evaluating network vulnerability and robustness of graphs.
\toolbox{} contains 22 graph robustness measures with both original and fast approximate versions when possible; 
17 failure and attack mechanisms; 15 heuristic and optimization based defense techniques; and 4 simulation tools.
\toolbox{} also supports a large number of GPU accelerated robustness measures.
To maintain the integrity of the code base, \toolbox{} uses continuous integration to run a suite of test cases on every commit.
To the best of our knowledge, this makes \toolbox{} the most comprehensive open-source framework for network robustness analysis to date.

\medskip
\noindent\textbf{2. Open-Source \& Permissive Licensing.}
Our goal is to democratize the tools needed to study network robustness; assisting researchers and practitioners in the analysis of their own networks.
As such, we open-source the \textit{code} on Github and PyPi with an \textit{MIT license} 
available at: \url{https://github.com/safreita1/TIGER}.

\smallskip
\noindent\textbf{3. Extensive Documentation \& Tutorials.}
We extensively document the functionality of \toolbox{}, providing docstrings for each function and class, along with quick examples on how to use the robustness measures, attacks, defenses, and simulation frameworks.
In addition, we provide $5$ detailed tutorials---one for every major component of \toolbox{}'s functionality--- on multiple large-scale, real-world networks, including \textit{every} figure and plot shown in this paper.
Users with Python familiarity will be able to readily pick up \toolbox{} for analysis with their own data.

\smallskip
\noindent\textbf{4. Community Impact.}
\toolbox{} helps enable reproducible research and the timely dissemination of new and current ideas in the area of network robustness and vulnerability analysis.
As part of the newly released Nvidia Data Science Teaching Kit, \toolbox{} will be used by educators and researchers across the world. 
\toolbox{} has been integrated into 
the Nvidia Data Science Teaching Kit available to educators across the world; 
and Georgia Tech’s Data and Visual Analytics with over 1,000 students.
Since this is a \textit{large} and \textit{highly active} field across many disciplines of science and engineering, we anticipate that \toolbox{} will have significant impact.
As the field grows, we will continue to update \toolbox{} with new techniques and features.

\section{TIGER Robustness Measures}\label{section:metrics}

\toolbox{} contains 22 robustness measures, grouped into one of three categories depending on whether the measure utilizes the graph, adjacency, or Laplacian matrix.
We present 3 representative robustness measures, one from each of the three categories, to extensively discuss.
For detailed description and discussion of all 22 measures, we refer the reader to the online documentation.

\smallskip
\noindent
\textbf{Terminology and Notation.} 
As the study of graphs has been carried out in a variety of fields (e.g., mathematics, physics, computer science), the terminology often varies from field to field. 
As such, we refer to the following word pairs interchangeably: (network, graph), (vertex, node), (edge, link).
Throughout the paper, we follow standard practice and use capital bold letters for matrices (e.g., $\bm{A}$), lower-case bold letters for vectors (e.g., $\bm{a}$). 
Also, we focus on undirected and unweighted graphs.

\begin{figure*}[t]
    \centering
    \includegraphics[width=\textwidth]{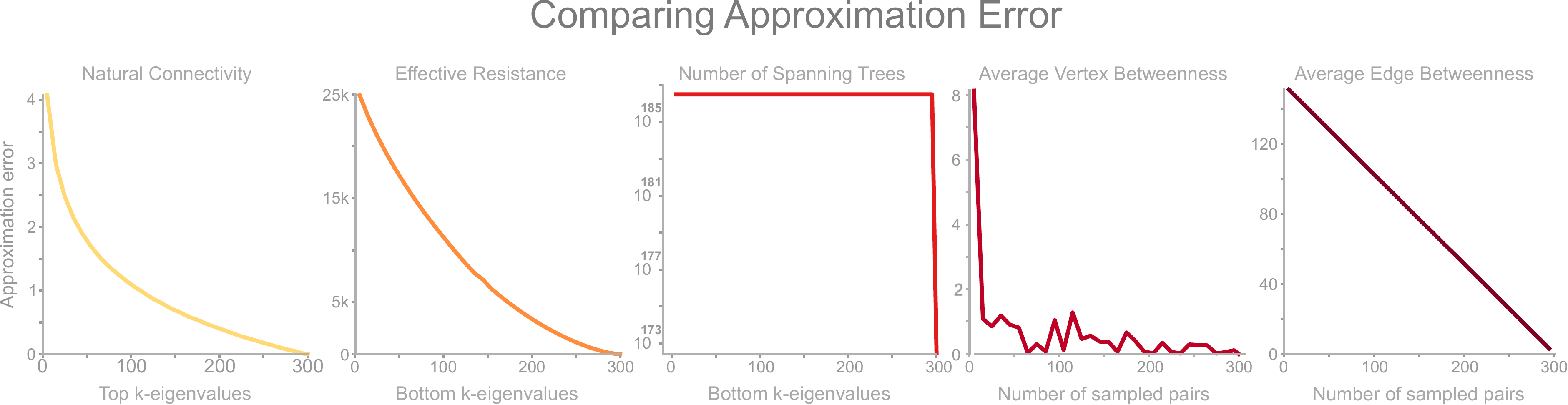}
    \caption{Error of 5 fast, approximate robustness measures supported by \toolbox{}.
    Parameter $k$ represents the trade-off between speed (low $k$) and precision (high $k$). 
    To measure approximation efficacy, we vary $k\in [5, 300]$ in increments of $10$ and measure the error between the approximate and original measure averaged over 30 runs on a clustered scale-free graph with 300 nodes.
    }
    \label{fig:measure_error}
\end{figure*}

\subsection{Example Measures}\label{subsec:measure_overview}

\noindent\textbf{Average vertex betweenness ($\bar{b}_v$)}\label{subsec:vertex-betweenness}
of a graph $G=(\altmathcal{V}, \altmathcal{E})$ is the summation of vertex betweenness $b_u$ for every node $u\in V$, where vertex betweenness for node $u$ is defined as the number of shortest paths that pass through $u$ out of the total possible shortest paths

\begin{equation}\label{eq:vertex-betweenness}
    \bar{b}_v = \sum_{u\in V}\sum_{s\in V}\sum_{\substack{t\in V \\ s\neq t\neq u}} \frac{n_{s,t}(u)}{n_{s,t}}
\end{equation}

where $n_{s, t}(u)$ is the number of shortest paths between $s$ and $t$ that pass through $u$ and $n_{s, t}$ is the total number of shortest paths between $s$ and $t$~\cite{freeman1977set}.
Average vertex betweenness has a natural connection to graph robustness since it measures the average load on vertices in the network.
The smaller the average the more robust the network, since load is more evenly distributed across nodes.

\smallskip
\noindent\textbf{Spectral scaling ($\xi$)} indicates if a network is simultaneously sparse and highly connected, known as ``good expansion'' (GE)~\cite{hoory2006expander,estrada2006network}.
Intuitively, we can think of a network with GE as a network lacking bridges or bottlenecks.
In order to determine if a network has GE, \cite{estrada2006network} proposes to combine the spectral gap measure with odd subgraph centrality $SC_{odd}$, which measures the number of odd length closed walks a node participates in.
Formally, \textit{spectral scaling} is described in Equation~\ref{eq:expansion_character},

\begin{equation}\label{eq:expansion_character}
    \xi(G) = \sqrt{\frac{1}{n} \sum_{i=1}^{n} \{log[\bm{u}_1(i)] - [log\bm{A} + \frac{1}{2} log[SC_{odd}(i)]] \}^2 }
\end{equation}

where $\bm{A}$ = $[sinh(\lambda_1)]^{-0.5}$, $n$ is the number of nodes, and $\bm{u}_1$ is the first eigenvector of adjacency matrix $\bm{A}$.
The closer $\xi$ is to zero, the better the expansion properties and the more robust the network.
Formally, a network is considered to have GE if $\xi < 10^{-2}$, the correlation coefficient $r < 0.999$ and the slope is $0.5$.

\smallskip
\noindent\textbf{Effective resistance ($R$)}
views a graph as an electrical circuit where an edge $(i, j)$ corresponds to a resister of $r_{ij}$ = $1$ Ohm and a node $i$ corresponds to a junction.
As such, the effective resistance between two vertices $i$ and $j$, denoted $R_{ij}$, is the electrical resistance measured across $i$ and $j$ when calculated using Kirchoff's circuit laws.
Extending this to the whole graph, we say the \textit{effective graph resistance} $R$ is the sum of resistances for all distinct pairs of vertices~\cite{ellens2013graph,ghosh2008minimizing}.
Klein and Randic~\cite{klein1993resistance} proved this can be calculated based on the sum of the inverse non-zero Laplacian eigenvalues:

\begin{equation}
    R = \frac{1}{2}\sum_{i, j}^{n} R_{ij} = n\sum_{i=2}^{n} \frac{1}{\mu_i}
\end{equation}

As a robustness measure, effective resistance measures how well connected a network is, where a smaller value indicates a more robust network~\cite{ghosh2008minimizing,ellens2013graph}. 
In addition, the effective resistance has many desirable properties, including the fact that it strictly decreases when adding edges, and takes into account both the number of paths between node pairs and their length~\cite{ellens2011effective}.

\subsection{Measure Implementation \& Evaluation}\label{subsec:measure_comparison}
Our goal for \toolbox{} is to implement each robustness measure in a clear and concise manner to facilitate code readability, while simultaneously optimizing for execution speed.
Each robustness measure is wrapped in a function that abstracts mathematical details away from the user;
and any default parameters are set for a balance of \textit{speed} and \textit{precision}.
Below we compare the efficacy of 5 fast, approximate robustness measures, followed by an analysis of the scalability of all 22 measures.

\smallskip
\noindent\textbf{Approximate Measures.}
It turns out that a large number of robustness measures have difficulty scaling to large graphs.
To help address this, we implement and compare 5 fast approximate measure, three spectral based (natural connectivity, number of spanning trees, effective resistance), and two graph based (average vertex betweenness, average edge betweenness)~\cite{chan2016optimizing,brandes2007centrality}.
To approximate \textit{natural connectivity} we use the top-$k$ eigenvalues of the adjacency matrix as a low rank approximation~\cite{chan2016optimizing,malliaros2012fast}.
For the \textit{number of spanning trees} and \textit{effective resistance} we take the bottom-$k$ eigenvalues of the Laplacian matrix~\cite{chan2016optimizing}.
For graph measures, \textit{average vertex betweenness} and \textit{average edge betweenness}, we randomly sample $k$ nodes to calculate centrality.
In both cases, the parameter $k$ represents the trade-off between speed (low $k$) and precision (high $k$). 
When $k$ is equal to the number of nodes $n$ in the graph, the approximate measure is equivalent to the original.
 
To determine the efficacy of each approximation measure we vary $k\in [5, 300]$ in increments of $10$, and measure the absolute error between the approximate and original measure, averaged over 30 runs on a clustered scale free graph containing 300 nodes.
In Figure~\ref{fig:measure_error}, we observe that average vertex betweenness accurately approximates the original measure using $\sim$10\% of the nodes in the graph. 
This results in a significant speed-up, and is in line with prior research~\cite{brandes2007centrality}.
While the absolute error for each spectral approximation is large, these approximations find utility in measuring the relative change in graph robustness after a series of perturbations (i.e., addition or removal of nodes/edges).
While not immediately obvious, this can enable the development a wide range of optimization based defense techniques~\cite{chan2016optimizing,chan2014make}.

\subsection{Running Robustness Measures in TIGER}
The code block in Listing 1 illustrates how \toolbox{} abstracts the code complexity away from the user, enabling them to quickly evaluate the robustness of their own network data in a simple manner.
In line 1, we import a helper function to generate various NetworkX graphs; line 2 imports a utility function to run the specified robustness measure; line 5 creates a Barabasi-Albert (BA) graph with 1000 nodes; and in lines 8 and 12 we calculate the graph's spectral radius and effective resistance, respectively.

\begin{lstlisting}[language=Python, caption=Measuring the spectral radius and effective resistance of a Barabasi-Albert (BA) graph using TIGER]
from graph_tiger.graphs import graph_loader
from graph_tiger.measures import run_measure

# Load a Barabasi-Albert graph with 1000 nodes
graph = graph_loader(graph_type='BA', n=1000, seed=1)

# Calculate graph's spectral radius
sr = run_measure(graph, measure='spectral_radius')
print("Spectral radius:", sr)

# Calculate graph's effective resistance
er = run_measure(graph, measure='effective_resistance')
print("Effective resistance:", er)
\end{lstlisting}

\begin{figure*}[t]
    \centering
    \includegraphics[width=0.95\textwidth]{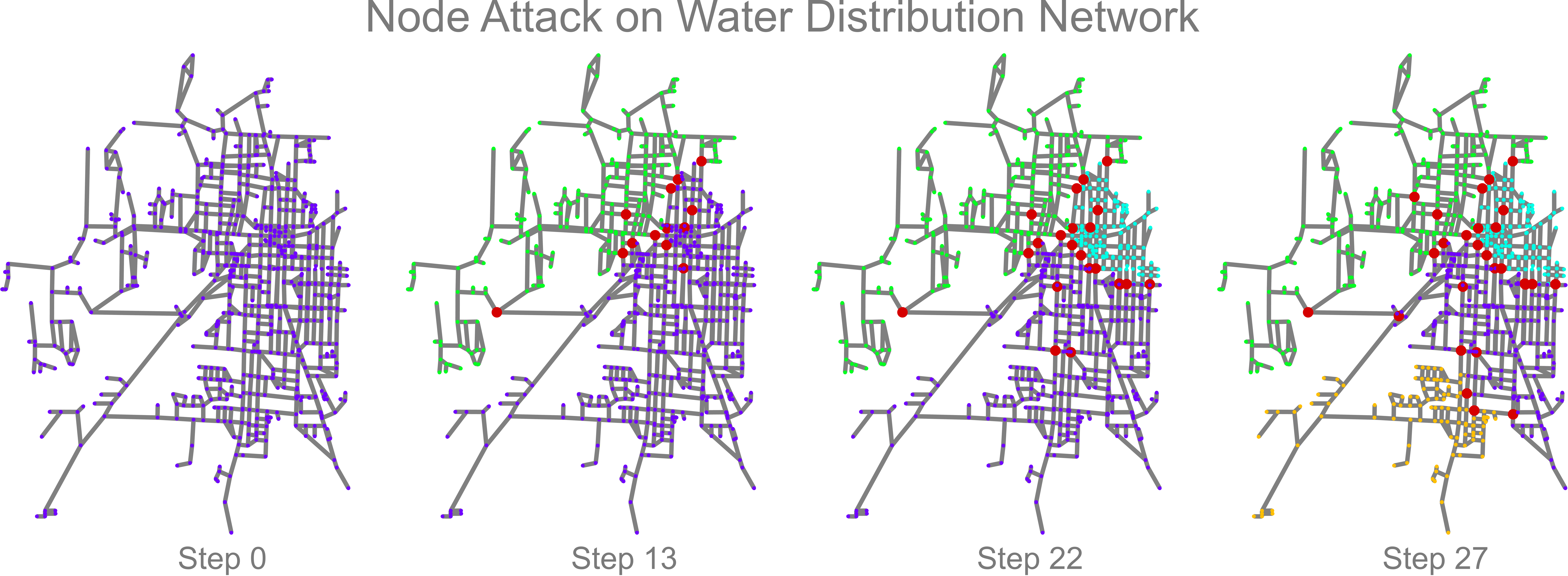}
    \caption{\toolbox{} simulation of an RD node attack on the KY-2 water distribution network. 
    Step 0: network starts under normal conditions; at each step a node is removed by the attacker (red nodes). 
    Step 13, 22 \& 27: after removing only a few of the 814 nodes, the network splits into two and three and four disconnected regions, respectively.}
    \label{fig:network_attack}
\end{figure*}

\section{TIGER Attacks}\label{section:attack}

There are two primary ways a network can become damaged---(1) \textit{natural failure} and (2) \textit{targeted attack}.
Natural failures typically occur when a piece of equipment breaks down from natural causes. 
In the study of graphs, this would correspond to the removal of a node or edge in the graph.
While random network failures regularly occur, they are typically less severe than targeted attacks.
This has been shown to be true across a range of graph structures~\cite{xia2010cascading,beygelzimer2005improving}.
In contrast, targeted attacks carefully select nodes and edges in the network for removal in order to maximally disrupt network functionality.
As such, we focus the majority of our attention to targeted attacks.
In Section~\ref{subsec:attack_overview}, we provide a high-level overview of several network failure and attack strategies.
Then, in Section~\ref{subsec:comparing_attacks} we highlight 10 attack strategies implemented in \toolbox{}.

\subsection{Attack Strategies}\label{subsec:attack_overview}
We showcase an example attack in Figure~\ref{fig:network_attack} on the Kentucky KY-2 water distribution network~\cite{hernadez2016water}.
The network starts under normal conditions (far left), and at each step an additional node is removed by the attacker (red nodes). 
After removing only 13 of the 814 nodes, the network is split into two separate regions. 
By step 27, the network splits into four disconnected regions.
In this simulation, and in general, attack strategies rely on node and edge centrality measures to identify candidates.
Below, we highlight several attack strategies~\cite{holme2002attack} contained in \toolbox.

\smallskip
\noindent\textbf{Initial degree removal (ID)} targets nodes with the highest degree $\delta_v$.
This has the effect of reducing the total number of edges in the network as fast as possible~\cite{holme2002attack}.
Since this attack only considers its neighbors when making a decision, it is considered a \textit{local attack}.
The benefit of this locality is low computational overhead.

\smallskip
\noindent\textbf{Initial betweenness removal (IB)} targets nodes with high betweenness centrality $b_v$.
This has the effect of destroying as many paths as possible~\cite{holme2002attack}.
Since path information is aggregated from across the network, this is considered a \textit{global attack} strategy.
Unfortunately, global information comes with significant computational overhead compared to a local attacks.

\begin{figure}[b]
    \centering
    \includegraphics[width=\linewidth]{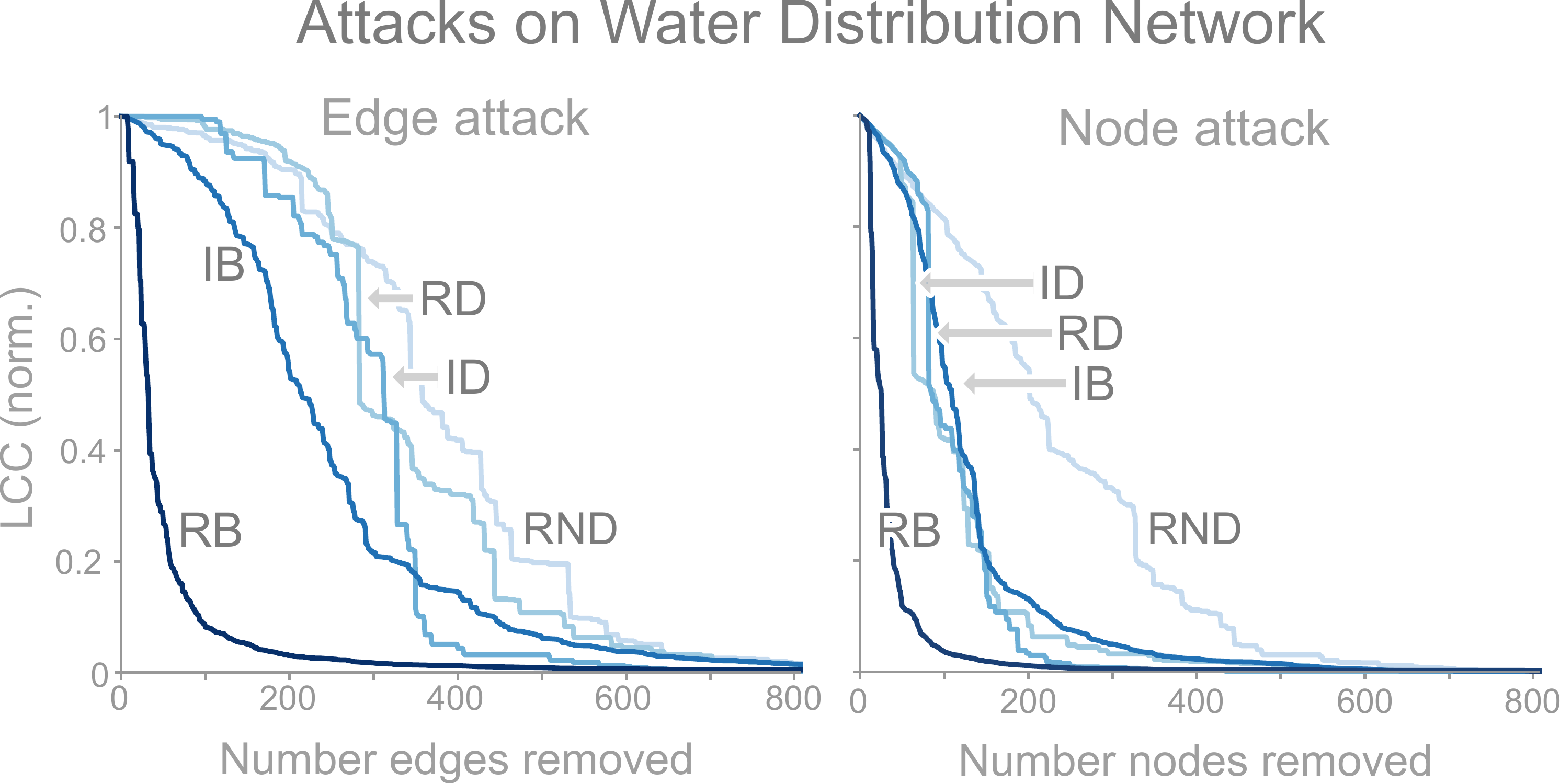}
    \caption{Efficacy of 5 edge attacks (left) and 5 node attacks (right) on the KY-2 water distribution network.
    The most effective attack (RB) disconnects approximately $50\%$ of the network with less than 30 removed edges (or nodes).}
    \label{fig:attack_plots}
\end{figure}

\smallskip
\noindent\textbf{Recalculated degree ($RD$)} and \textbf{betweenness removal ($RB$)} follow the same process as $ID$ and $IB$, respectively, with one additional step to recalculate the degree (or betweenness) distribution after a node is removed.
This recalculation often results in a stronger attack, 
however, recalculating these distributions adds a significant amount of computational overhead to the attack.

\subsection{Comparing Strategies}\label{subsec:comparing_attacks}
To help \toolbox{} users determine the effectiveness of attack strategies, we evaluate $5$ node and $5$ edge attacks on the Kentucky KY-2 water distribution network in Figure~\ref{fig:attack_plots}.
We begin by analyzing each node attack strategy---$ID$, $RD$, $IB$, $RB$ and $RND$ (random selection)---on the left-side of Figure~\ref{fig:attack_plots}. 
Attack success is measured based on how fractured the network becomes when removing nodes from the network. 
We identify three key observations---(i) random node removal ($RND$) is not an effective strategy on this network structure; (ii) $RB$ is the most effective attack strategy; and (iii) the remaining three attacks are roughly equivalent, falling somewhere between $RND$ and $RB$.

Analyzing Figure~\ref{fig:network_attack}, we can gain insight into why $RB$ is the most effective of the attacks.
If we look carefully, we observe that certain nodes (and edges) in the network act as key bridges between various network regions.
As a result, attacks able to identify these bridges are highly effective in disrupting this network.
In contrast, degree based attacks are less effective, likely due to the balanced degree distribution.
The analysis is similar for edge based attacks.

\subsection{Running Network Attacks in TIGER}
The code block in Listing 2 illustrates how \toolbox{} users can quickly run a network attack by modifying 3 parameters---(1) the number of attack simulations `runs', (2) the number of nodes to remove `steps', and (3) the attack strategy `attack'. 
The output of the simulation is a plot of graph robustness (e.g., largest connected component by default) versus attack strength.

\begin{lstlisting}[language=Python, caption=Attacking the Kentucky KY-2 water distribution network using TIGER]
from graph_tiger.attacks import Attack
from graph_tiger.graphs import graph_loader

params = {
     'runs': 1,            # number of simulations
     'steps': 30,          # remove 1 node per step
     'attack': 'rd_node',  # specify attack
     'seed': 1,            # reproducibility
}
 
# Load Kentucky KY-2 water distribution network
graph = graph_loader(graph_type='ky2')

# Run and plot attack simulation
a = Attack(graph, **params)
results = a.run_simulation() 
a.plot_results(results)
\end{lstlisting}

\section{TIGER Defenses}\label{sec:defenses}
The same centrality measures effective in attacking a network are important to network defense (e.g., degree, betweenness, PageRank, eigenvector, etc.).
In fact, if an attack strategy is known a priori, node monitoring can largely prevent an attack altogether.
In Section~\ref{subsec:defense_overview}, we provide a high-level overview of several heuristic and optimization based defense techniques.
Then, in Section~\ref{subsec:defense_comparison} we show \toolbox{} users how several defense techniques can be used to robustify an attacked network.

\subsection{Defense Strategies}\label{subsec:defense_overview}
We categorize defense techniques based on whether they operate heuristically, modifying graph structure independent of a robustness measure, or by optimizing for a particular robustness measure~\cite{chan2016optimizing}.
Within each  category a network can be defended i.e., improve its robustness by---(1) \textit{edge rewiring}, (2) \textit{edge addition}, or (iii) \textit{node monitoring}. 
Edge rewiring is considered a \textit{low} cost, \textit{less} effective version of edge addition. 
On the other hand, edge addition almost always provides stronger defense~\cite{beygelzimer2005improving}.
Node monitoring provides an orthogonal mechanism to increase network robustness by monitoring (or removing) nodes in the graph.
This has an array of applications, including: (i) preventing targeted attacks, (ii) mitigating cascading failures, and (iii) reducing the spread of network entities.
Below, we highlight several heuristic and optimization based techniques contained in \toolbox.

\smallskip
\noindent\textbf{Heuristic Defenses.}
We overview $5$ edge rewiring and addition defenses~\cite{beygelzimer2005improving}, and compare the effectiveness of them in
Section~\ref{subsec:defense_comparison}:

\begin{enumerate}[label=\arabic*., topsep=4pt, leftmargin=*, itemsep=3pt]
    \item \textit{Random addition}: adds an edge between two random nodes.
    
    \item \textit{Preferential addition}: adds an edge connecting two nodes with the lowest degrees.
    
    \item \textit{Random edge rewiring}: removes a random edge and adds one using (1).
    
    \item \textit{Random neighbor rewiring}: randomly selects neighbor of a node and removes the edge. An edge is then added using (1).
    
    \item \textit{Preferential random edge rewiring}: selects an edge, disconnects the higher degree node, and reconnects to a random one.
\end{enumerate}

\smallskip
\noindent\textbf{Optimization Defenses.}
We discuss the Netshield node monitoring technique which identifies key nodes in a network to reduce the spread of entity dissemination (e.g., viruses)~\cite{tong2010vulnerability}.
To minimize the spread of entities, Netshield minimizes the spectral radius of the graph $\lambda_1$ by selecting the best set $S$ of $k$ nodes to remove from the graph (i.e., maximize eigendrop). 
In order to evaluate the goodness of a node set $S$ for removal, \cite{tong2010vulnerability} proposes the Shield-value measure:

\begin{equation}
    Sv(S) = \sum_{i\in S}2\lambda_1 \bm{u}_1(i)^2 - \sum_{i,j\in S} A(i,j)\bm{u}(i)\bm{u}(j)
\end{equation}

\noindent The intuition behind this equation is to select nodes for monitoring that have high eigenvector centrality (first term), while penalizing neighboring nodes to prevent grouping (second term).
We demonstrate the utility of this defense mechanism in Section~\ref{sec:simulations}.

\begin{figure}[t]
    \centering
    \includegraphics[width=0.95\linewidth]{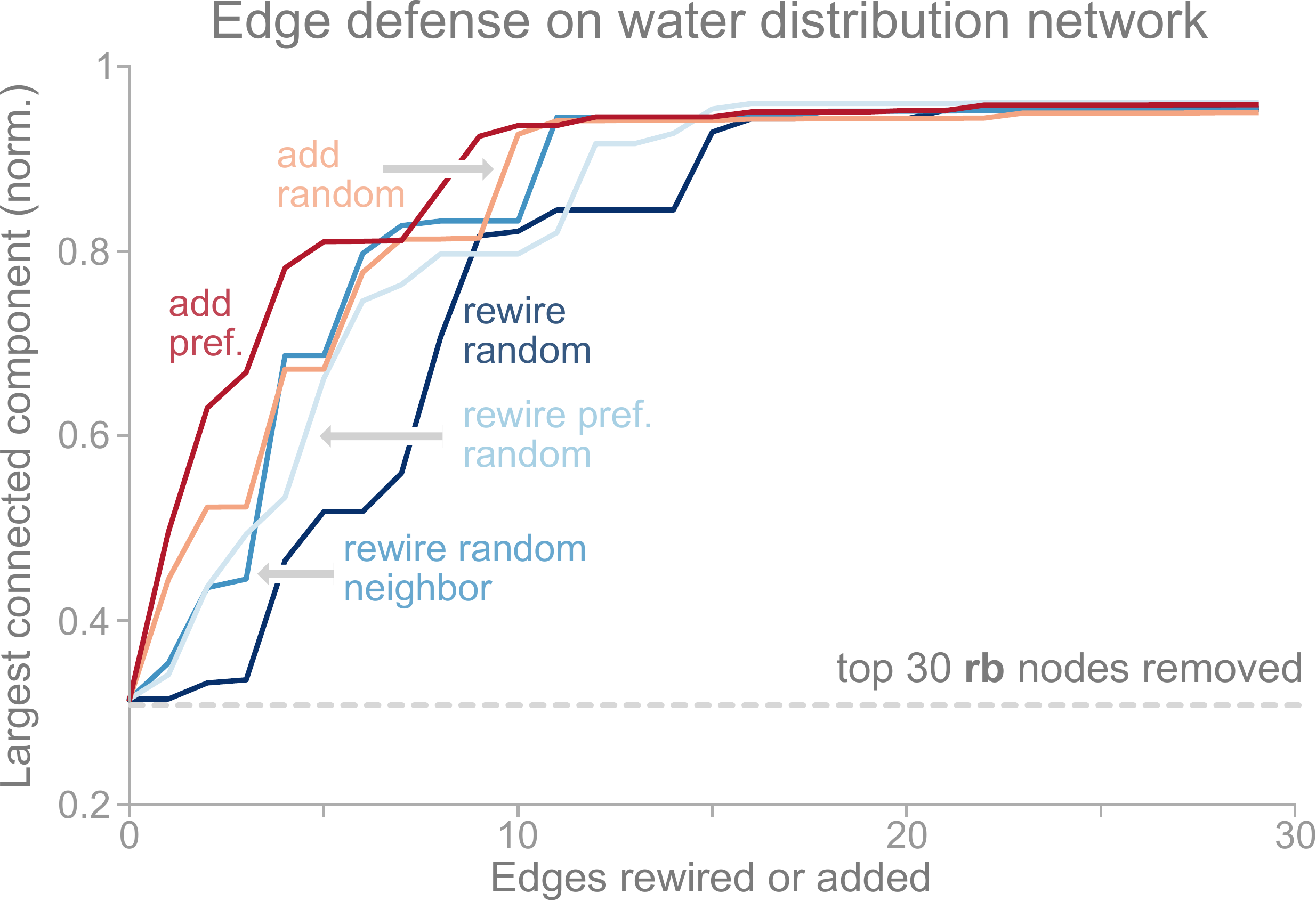}
    \caption{Comparing ability of 5 edge defenses to improve KY-2 network robustness after removing 30 nodes via RB attack.
    Edge addition performs the best, with random edge rewiring performing the worst.}
    \label{fig:edge_defense}
\end{figure}

\subsection{Comparing Strategies}\label{subsec:defense_comparison}
To help users evaluate the effectiveness of defense techniques, we compare 5 edge defenses on the Kentucky KY-2 water distribution network, averaged over 10 runs, in Figure~\ref{fig:edge_defense}.
The network is initially attacked using the $RB$ attack strategy (30 nodes removed), and the success of each defense is measured based on how it can reconnect the network by adding or rewiring edges in the network (higher is better). 
Based on Figure~\ref{fig:edge_defense}, we identify three key observations---(i) preferential edge addition performs the best; (ii) edge addition generally outperforms rewiring strategies; and (iii) random neighbor rewiring typically performs better than the other rewiring strategies.

\subsection{Running Network Defenses in TIGER}
The code block in Listing 3 illustrates how \toolbox{} users can quickly setup a network defense simulation.
There are 6 core parameters the user needs to set---the number of defense simulations `runs', the defense strategy `defense', the number of edges to add or rewire `steps', the attack strategy `attack', and the number of nodes or edges to remove 'k\_a'.
The output of the simulation is a plot showing the ability of the network to recover after it has been attacked.

\begin{lstlisting}[language=Python, caption=Defending the Kentucky KY-2 water distribution network using TIGER]
from graph_tiger.defenses import Defense
from graph_tiger.graphs import graph_loader

params = {
     'runs': 1,            # number of simulations
     'steps': 30,          # rewire 1 edge per step
     'defense': 'rewire_edge_preferential',
     'attack': 'rd_node',  # attack strategy
     'k_a': 30,            # attack strength
}
 
# Load Kentucky KY-2 water distribution graph
graph = graph_loader(graph_type='ky2')

# Run and plot defense simulation
d = Defense(graph, **params)
results = d.run_simulation() 
d.plot_results(results)
\end{lstlisting}
\section{TIGER Simulation Tools}\label{sec:simulations}
We implement 4 broad and important types of robustness simulation tools~\cite{kermack1927contribution,watts1998collective,holme2002attack,beygelzimer2005improving,tong2010vulnerability}---(1) dissemination of network entities, 
(2) cascading failures 
(3) network attacks, see Section~\ref{section:attack}, and 
(4) network defense, see Section~\ref{sec:defenses}.
In Section~\ref{subsec:dissemination}, we discuss the implementation of an infectious disease models and how defense techniques implemented in \toolbox{} can be used to either \textit{minimize} or \textit{maximize} the network diffusion.
Then, in Section~\ref{section:attack}, we discuss the implementation of the cascading failure model and its interactions with \toolbox{} defense and attack strategies.

\begin{figure}[b]
    \centering
    \includegraphics[width=0.95\linewidth]{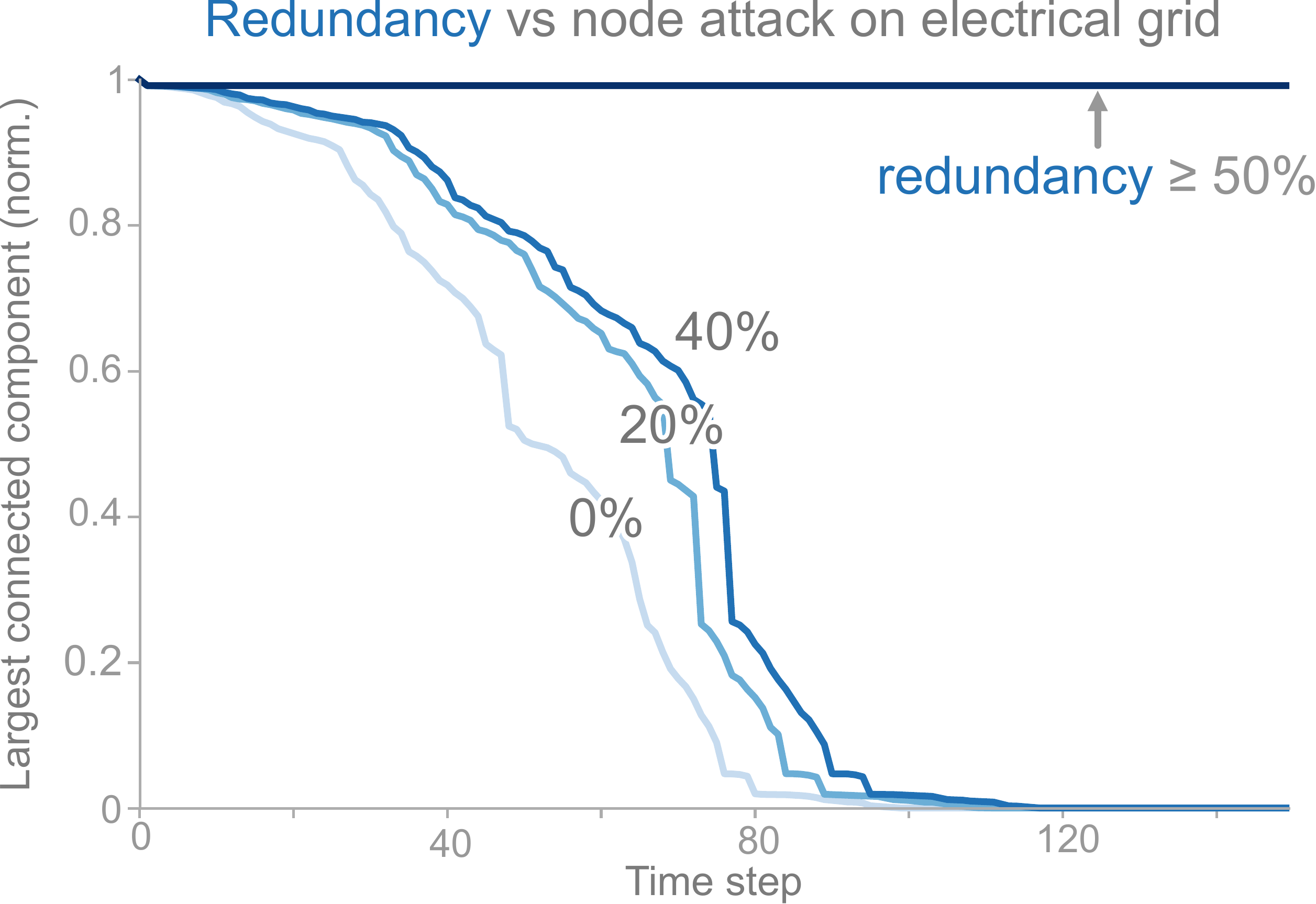}
    \vspace{-0.25cm}
    \caption{Effect of network redundancy $r$ on the US power grid where 4 nodes are overloaded using ID. 
    When $r\geq 50\%$ the network is able to redistribute the increased load.
    }
    \label{fig:cascading_plot}
    \vspace{-0.5cm}
\end{figure}

\begin{figure*}[t]
    \centering
    \includegraphics[width=0.95\textwidth]{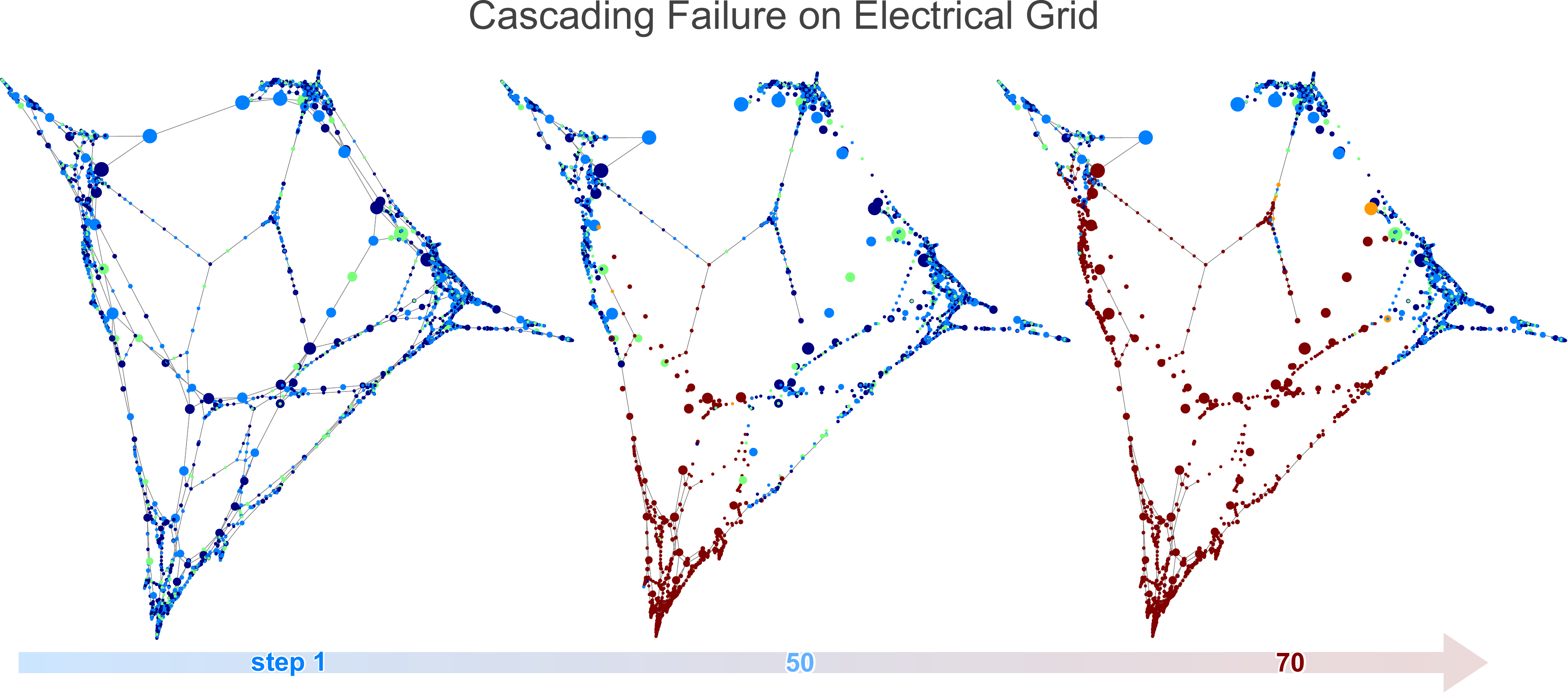}
    \caption{
    \toolbox{} cascading failure simulation on the US power grid network when 4 nodes are overloaded according to the ID attack strategy.
    Time step 1: shows the network under normal conditions.
    Time step 50: we observe a series of failures originating from the bottom of the network.
    Time step 70: most of the network has collapsed.}
    \label{fig:cascading_network}
\end{figure*}

\subsection{Cascading Failures}\label{subsec:cascading_failures}
Cascading failures often arise as a result of natural failures or targeted attacks in a network.
Consider an electrical grid where a central substation goes offline. 
In order to maintain the distribution of power, neighboring substations have to increase production in order to meet demand.
However, if this is not possible, the neighboring substation fails, which in turn causes additional neighboring substations to fail. 
The end result is a series of cascading failures i.e., a blackout~\cite{crucitti2004model}.
While cascading failures can occur in a variety of network types e.g., water, electrical, communication, we focus on the electrical grid.  
Below, we discuss the design and implementation of the cascading failure model and how \toolbox{} can be used to both \textit{induce} and \textit{prevent} cascading failures using the attack and defense mechanisms discussed in Sections~\ref{section:attack} and \ref{sec:defenses}, respectively.

\smallskip
\noindent\textbf{Design and Implementation.}
There are 3 main processes governing the network simulation---%
(1) \textit{capacity} of each node $c_v\in [0,1]$; 
(2) \textit{load} of each node $l_v\in U(0, l_{max})$; and 
(3) network \textit{redundancy} $r\in [0, 1]$.
The capacity of each node $c_v$ is the the maximum load a node can handle, which is set based on the node's normalized betweenness centrality~\cite{hernandez2013probabilistic}.
The load of each node $l_v$ represents the fraction of maximum capacity $c_v$ that the node operates at.
Load for each node $c_v$ is set by uniformly drawing from $U(0, l_{max})$, where $l_{max}$ is the maximum initial load.
Network redundancy $r$ represents the amount of reserve capacity present in the network i.e., auxiliary support systems.
At the beginning of the simulation, we allow the user to attack and defend the network according to the node attack and defense strategies in Sections~\ref{section:attack} and \ref{sec:defenses}, respectively.
When a node is attacked it becomes ``overloaded'', causing it to fail and requiring the load be distributed to the neighbors.
When defending a node we increase it's capacity to protect against attacks.

\smallskip
\noindent\textbf{Simulating cascading failures.}
To help users visualize cascading failures induced by targeted attacks, we enable them to create visuals like Figure~\ref{fig:cascading_network}, where
we overload 4 nodes selected by the ID attack strategy on the US power grid dataset~\cite{watts1998collective} ($l_{max}=0.8$).
Node size represents capacity i.e., larger size $\rightarrow$ higher capacity, and color indicates the load of each node on a gradient scale from blue (low load) to red (high load); dark red indicates node failure (overloaded).
Time step 1 shows the network under normal conditions; at step 50 we observe a series of failures originating from the bottom of the network; by step 70 most of the network has collapsed.
To assist users in summarizing simulation results over many configurations, we enable them to create plots like Figure~\ref{fig:cascading_plot}, which shows the effect of network redundancy $r$ when 4 nodes are overloaded by the ID attack strategy. 
At $50\%$ redundancy, we observe a critical threshold where the network is able to redistribute the increased load.
For $r<50\%$, the cascading failure can be delayed but not prevented.

\subsection{Running Cascading Failures in TIGER}
The code block in Listing 4 shows how \toolbox{} users can quickly setup a cascading failure simulation.
There are 3 simulation specific parameters---the max node lode `l', node redundancy `r', and maximum node capacity 'c' (based on betweenness centrality).
We set the attack and defense parameters, similar to Listings 2 and 3, respectively. 
The simulation output is a plot measuring the `health' or robustness of the network over time.
Users can optionally generate image snapshots and a video simulation of the cascading failure on the network data.

\begin{lstlisting}[language=Python, caption=Cascading failure simulation on U.S. electrical grid using TIGER]
from graph_tiger.cascading import Cascading
from graph_tiger.graphs import graph_loader

params = {
   'runs': 1,            # number of simulations
   'steps': 100,         # simulation time steps
   'l': 0.8,             # max node load
   'r': 0.2,             # node redundancy
   'c': int(0.1 * len(graph)),  # node capacity approx.
   
   'robust_measure': 'largest_connected_component',
   'k_a': 30,            # attack strength
   'attack': 'rd_node',  # attack strategy
   'k_d': 0,             # defense strength
   'defense': None,      # defense strategy
}

# Load U.S. electrical grid graph
graph = graph_loader('electrical')

# Run and plot cascading failure simulation
cascading = Cascading(graph, **params)
results = cascading.run_simulation()
cascading.plot_results(results)

\end{lstlisting}

\subsection{Dissemination of Network Entities}\label{subsec:dissemination}
A critical concept in entity dissemination is \textit{network diffusion}, which attempts to capture the underlying mechanism enabling network propagation.
In order to augment this diffusion process, \toolbox{} leverages the defense techniques in Section~\ref{sec:defenses} for use with two prominent diffusion models: the flu-like susceptible-infected-susceptible (SIS) model, and the vaccinated-like susceptible-infected-recovered (SIR) model~\cite{kermack1927contribution}.
For example, to \textit{minimize} the ability of viruses to spread we can monitor (remove) nodes in the graph to reduce graph connectivity.
On the other hand, if want to \textit{maximize} network diffusion e.g., marketing campaign, we can use defense techniques like edge rewiring or addition to increase graph connectivity.
Below, we highlight the SIS infectious disease model and how \toolbox's defense techniques can help contain a simulated outbreak.

\smallskip
\noindent\textbf{Design and Implementation.}
Each node in the SIS model can be in one of two states, infected $I$ or susceptible $S$. 
At each time step $t$, an infected node $v$ has a probability $\beta$ of infecting each of it's uninfected neighbors $u\in N(v)$.
After this, each infected node $v$ has a probability $\delta$ of healing and becoming susceptible again.
The relationship between the birth rate $\beta$, death rate $\delta$ and the spectral radius $\lambda_1$ of the graph has been a widely studied topic.
In \cite{wang2003epidemic}, they show that the spectral radius of a graph is closely tied to the epidemic threshold $\tau$ of a network in an SIS model.
In particular, they prove that $\frac{\beta}{\delta} < \tau=\frac{1}{\lambda_{1}}$. 
This means for a given virus strength $s$, an epidemic is more likely to occur on a graph with larger $\lambda_1$.
As such, we say that a virus has an effective strength $s = \lambda_1 \cdot b / d$, where a larger $s$ means a stronger virus~\cite{tong2010vulnerability}.

\begin{figure}[t]
    \centering
    \includegraphics[width=\linewidth]{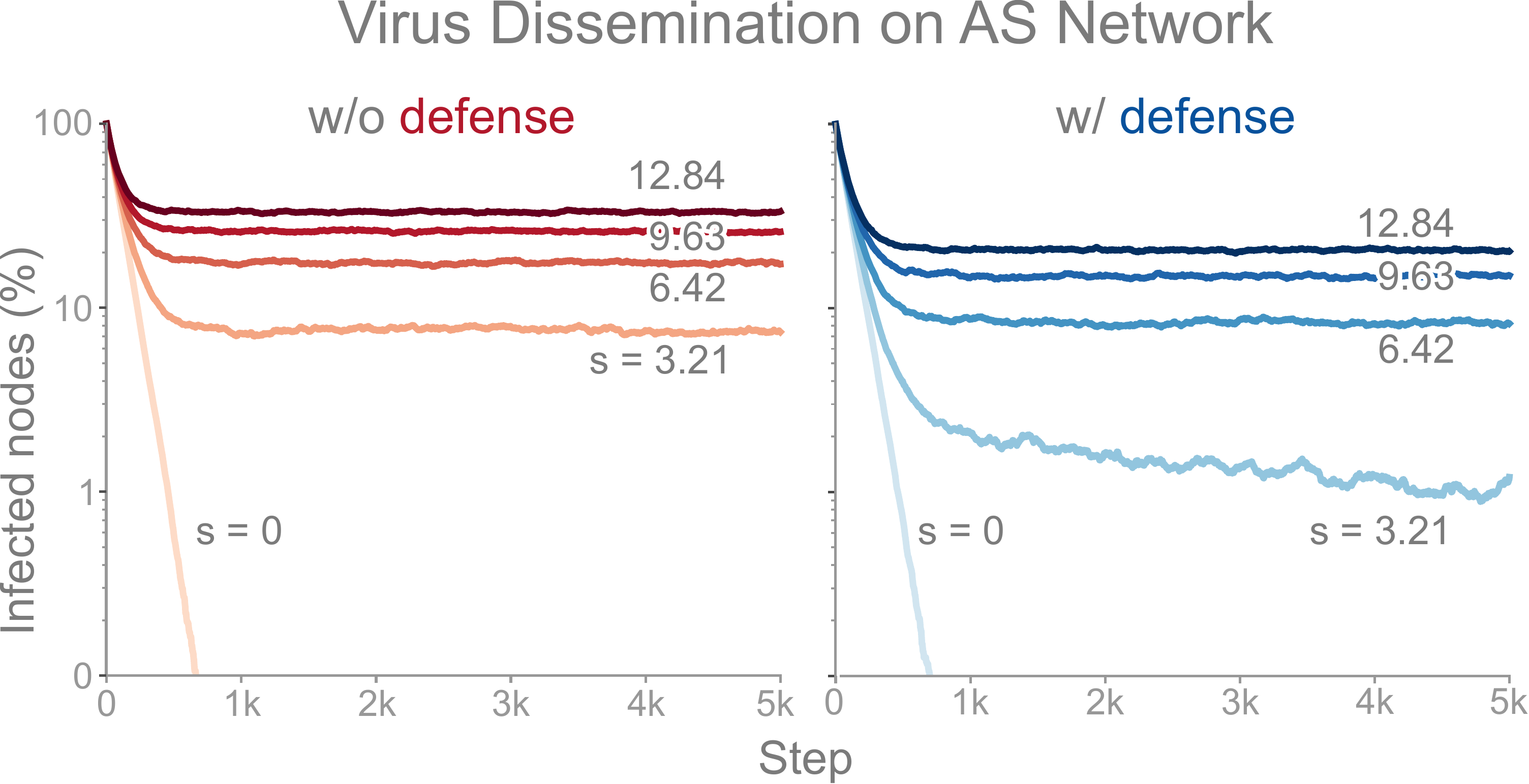}
    \caption{SIS simulation with 5 virus strengths on the Oregon-1 Autonomous System network. No defense (left), Netshield defense (right).}
    \label{fig:virus_plots}
\end{figure}

\smallskip
\noindent\textbf{Simulating dissemination of entities.}
To help users visualize the dissemination process, we enable them to create visuals like Figure~\ref{fig:virus_simulation}, where we run an SIS computer virus simulation ($s=3.21$) on the Oregon-1 Autonomous System network~\cite{leskovec2005graphs}.
The top of Figure~\ref{fig:virus_simulation} shows the virus progression when defending 5 nodes selected by Netshield~\cite{tong2010vulnerability}.
By time step 1000, the virus has nearly died out.
The bottom of Figure~\ref{fig:virus_simulation} shows that the virus remains endemic without defense.
To assist users in summarizing model results over many configurations, we enable them to create plots like Figure~\ref{fig:virus_plots}, which show results for 5
SIS effective 
virus strengths 
$s=\{0, 3.21, 6.42, 9.63, 12.84\}$
over a period of 5000 steps.

\subsection{Running Entity Dissemination in TIGER}
The code block in Listing 5 shows how \toolbox{} users can run an entity dissemination simulation by setting a few key parameters---the type of entity simulation `model' (e.g., SIS, SIR), the virus birth rate `b', the virus death rate `d', and the fraction of the network that starts off infected `c'.
The simulation output is a plot of network infection over time.
In addition, users can optionally generate image snapshots and a video simulation of the entity dissemination on the network.

\begin{lstlisting}[language=Python, caption=Entity dissemination simulation on Oregon-1 Autonomous System network using TIGER]
from graph_tiger.diffusion import Diffusion
from graph_tiger.graphs import graph_loader

sis_params = {
   'runs': 1,      # number of simulations
   'steps': 5000,  # simulation time steps

   'model': 'SIS',
   'b': 0.00208,  # virus birth rate 
   'd': 0.01,     # virus death rate
   'c': 0.3,      # network % starting infected 
}

# Load Oregon-1 Autonomous System graph
graph = graph_loader('as_733')

# Run and plot entity dissemination simulation
diffusion = Diffusion(graph, **sis_params)
results = diffusion.run_simulation()

diffusion.plot_results(results)

\end{lstlisting}

\section{Conclusion}\label{sec:conclusion}
The study of network robustness is a critical tool in the characterization and understanding of complex interconnected systems.
Through analyzing and understanding the robustness of these networks we can: 
(1) quantify network vulnerability and robustness, 
(2) augment a network's structure to resist attacks and recover from failure, and 
(3) control the dissemination of entities on the network (e.g., viruses, propaganda). 
While significant research has been conducted on all of these tasks, no comprehensive open-source toolbox currently exists to assist researchers and practitioners in this important topic.
This lack of available tools hinders reproducibility and examination of existing work, development of new research, and dissemination of new ideas.
To address these challenges, we contribute \toolbox{}, an open-sourced Python toolbox containing 22 graph robustness measures with both original and fast approximate versions;
17 failure and attack strategies; 15 heuristic and optimization based defense techniques; and 4 simulation tools.
\toolbox{} is open-sourced at: \url{https://github.com/safreita1/TIGER}. 

\section{Acknowledgements}
This work was in part supported by the NSF grant IIS-1563816, GRFP (DGE-1650044), and a Raytheon fellowship.

\bibliographystyle{ACM-Reference-Format}
\bibliography{main}

\end{document}